\documentclass[
    ,final            
  ]
  {aipproc}

\layoutstyle{8x11single}

\newcommand{\gp}{$g_P^{}$}
\newcommand{\LS}{$\Lambda_S^{}$}

\begin{document}

\title{Muon Capture on the Proton}

\classification{25.30.Mr, 14.20.Dh}
\keywords      {muon capture, axial current, pseudoscalar form factor, muon lifetime}

\author{P. Winter\footnote{Representing the MuCap collaboration}}{
  address={Department of Physics, University of Washington, Seattle, WA 98195, USA}
}

\begin{abstract}
The MuCap experiment measures the singlet rate \LS\ of muon capture on the proton. A negative muon beam is stopped in a time projection chamber filled with ultra-pure hydrogen gas at 10 bar and room temperature. In combination with the surrounding decay electron detectors, the lifetime of muons in hydrogen can be measured to determine \LS\ to a final precision of 1\%. The capture rate is then used to derive the nucleon's pseudoscalar form factor \gp. Our first-stage result, \gp$= 7.3 \pm 1.1$ \cite{Andreev:2007wg}, will soon be updated with the final analysis of the full statistics reducing the error by a factor of $\sim 2$.
\end{abstract}

\maketitle


\section{Motivation}
The basic electro-weak process $\mu^- + p \rightarrow n + \nu$ provides an important probe to study the helicity structure of the weak interaction. The basic matrix element in this low-energy regime is given by an effective four-fermion coupling:
\[
\mathcal{M} \sim G_F^{} V_{ud}^{} \cdot \bar{\psi}_\nu \gamma_\alpha (1-\gamma_5) \psi_\mu \cdot \bar{\psi}_n (V^\alpha_{} - A^\alpha_{}) \psi_p,
\]
with a pure V-A structure of the leptonic current $L^{}_\mu = \bar{\psi}_\nu \gamma_\alpha (1-\gamma_5) \psi_\mu$. The hadronic current $J_\mu^{} =  \bar{\psi}_n (V^\alpha_{} - A^\alpha_{}) \psi_p$ also follows an underlying V-A structure which is dressed due to the substructure of the nucleon. Generally, both the vector $V^\alpha_{}$ and axial-vector $A^\alpha_{}$ parts can be written in terms of three form factors each. However, the two second-class currents can be neglected because symmetry principles require them to be small. Hence, the capture process involves $g_V(q_0^2)$, $g_M(q_0^2)$,  $g_A(q_0^2)$, and $g_P(q_0^2)$, i.e. the vector, magnetic, axial, and pseudoscalar form factors, respectively. The relevant momentum transfer is $q_0^2 = -0.88 m_\mu^2$. Given the experimentally well established quantities $g_V(q_0^2)$, $g_M(q_0^2)$, and $g_A(q_0^2)$, the least known of these form factors, $g_P(q_0^2)$, can be extracted from a measurement of \LS. At the same time, $g_P$ is precisely calculable within heavy baryon chiral perturbation theory with its underlying concept of chiral symmetry breaking. Thus a measurement of the predicted value $g_P^{} = 8.26 \pm 0.23$ \cite{Bernard:1994wn, Kaiser:2003dr} constitutes an important test of QCD symmetries

The experimental efforts in measuring ordinary muon capture (OMC) as well as radiative muon capture (RMC) span a long history. Figure \ref{fig-mucap} shows the two most precise results \cite{Bardin:1980mi, Wright:1998gi} before MuCap. Due to the details of molecular pp$\mu$ formation described in reviews \cite{Kammel:2010, Gorringe:2002xx}, these experiments exhibit a strong dependence on the poorly known ortho-para transition rate $\lambda_{op}$. The experimental conditions of MuCap were chosen to strongly suppress any dependence on $\lambda_{op}$. This can be seen in the plot from our first-stage result \cite{Andreev:2007wg} which is in agreement with the prediction from chiral perturbation theory (ChPT). Our final result with ten times more statistics will determine the singlet capture \LS\ at a precision of 1\% resulting in $g_P(q_0^2)$ with 7\% precision. 

\begin{figure}[htb]
\centering{\includegraphics[width=0.6\textwidth]{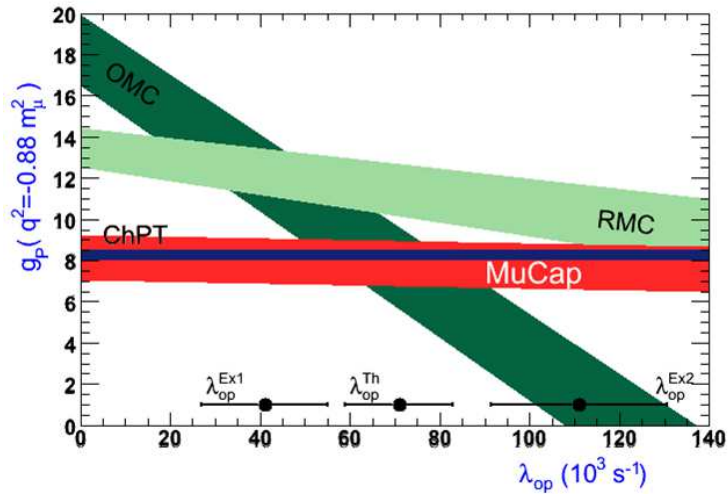}}
\caption{Previous most precise results from OMC \cite{Bardin:1980mi} and RMC \cite{Wright:1998gi} compared to the ChPT prediction \cite{Bernard:1994wn, Kaiser:2003dr} for \gp. Both experiments significantly depend on the poorly known molecular rate $\lambda_{op}$ ($\lambda_{op}^{Ex1}$ \cite{Bardin:1981cq}, $\lambda_{op}^{Ex2}$ \cite{Clark:2005as}, $\lambda_{op}^{Th}$ \cite{Bakalov:1980fm}). The MuCap result \cite{Andreev:2007wg} does not suffer from this dependence due to the chosen experimental conditions.\label{fig-mucapresults}}
\end{figure}

\section{The experiment and status of the analysis}
\begin{figure}[htb]
\centering{\includegraphics[width=0.55\textwidth]{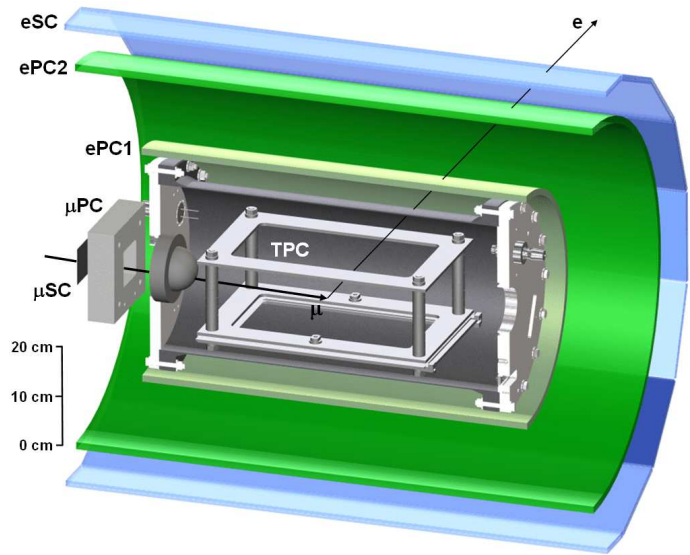}}
\caption{Simplified experimental setup of the MuCap detector.\label{fig-mucap}}
\end{figure}

The MuCap experiment located at the Paul Scherrer Institute (PSI), Switzerland, is schematically shown in figure \ref{fig-mucap}. Muons produced in the accelerator at the proton target are guided through the secondary beamline $\pi$E3 to enter the detector. The arriving muons are first registered in a scintillator ($\mu$SC) giving the precise start time for the lifetime measurement. The $\mu$SC's signal also triggers an upstream electrostatic kicker to deflect the muon beam for the following 25$\mu$s. In addition a two plane wire chamber ($\mu$PC) enables us to study the beam profile. Furthermore, the $\mu$PC and $\mu$SC signals are combined to disregard events with a second muon entering the detector during the 25$\mu$s after the first muon. This pileup happens because of the finite beam extinction of the kicker. At the center of the experiment is a time projection chamber (TPC) \cite{Egger:2011zz} filled with 10\,bar of ultra-clean hydrogen gas in which the incoming muon is stopped and forms a $\mu$p singlet atom. A high chemical purity of the hydrogen was achieved by constant circulation of the gas through an external cleaning unit \cite{Ganzha:2007uk}, reducing chemical contaminants to concentrations of less than 10\,ppb. Isotopic purity of less than 6\,ppb deuterium contamination was feasible by distilling the hydrogen gas in an isotope separation column prior to filling the TPC. Both isotopic and chemical purity are crucial to eliminate distortions to \LS\ arising from capture on non-hydrogen atoms as well as the time-dependent diffusion of $\mu$d atoms.

The ionization signals from the slowing muon in the TPC enable a full 3-dimensional reconstruction of the muon's trajectory and are key for selecting muons stopping in the hydrogen gas far away from non-hydrogen wall materials. The decay electrons' trajectories are then reconstructed in a set of two cylindrical wire chambers (ePC1 and ePC2) surrounding the TPC. The segmented scintillator hodoscope (eSC) provides the fast timing signal for the lifetime which is histogrammed from the difference of the eSC and $\mu$SC times. A fit of this histogram with an exponential and a flat background term to account for beam and cosmic accidentals yields the result of the lifetime of negative muons in hydrogen. The capture rate is then derived from the difference of the extracted lifetime to the one of positive muons \cite{Chitwood:2007pa, Barczyk:2007hp, Webber:2010zf} which cannot undergo the capture process. 

The analysis includes many consistency checks to ensure that the result is independent of the choice of various applied cuts. Currently we are in the process of finalizing the analysis of our full statistics of $\sim 1.5\cdot10^{10}$ negative muon decays which will improve the precision in \LS\ by a factor of $\sim 2$ compared to our first-stage result \cite{Andreev:2007wg}. Table \ref{tab:errors} shows the systematic and statistical errors for the 2007 result and our projected final uncertainties. Some of the systematic errors in the third column are already final, but a few need some more analysis to be completed. Once the collaboration settles on these outstanding issues, we can unblind the hidden offset that had been applied to our clock frequency during the measurement periods to prevent any analysis bias. In February 2011, we achieved an important intermediate step. Initially the $\sim 1.5\cdot10^{10}$ decays were collected in two running cycles with some differences in experimental conditions. These separate datasets had individual clock offsets and were analyzed independently. A person outside the collaboration provided conversion factors to bring the two datasets into a new common blinded frequency space allowing for the comparison of the two results. The agreement was excellent and therefore the collaboration is planning to reveal the final offset within the next 3 months.

\begin{table}[htb]
 \begin{tabular}{lcc}
 \hline
 \tablehead{1}{l}{b}{Source}
   & \tablehead{1}{c}{b}{2007 uncertainty [s$^{-1}$]}
   & \tablehead{1}{c}{b}{Projected final [s$^{-1}$]} \\
  \hline
Z$>$1 impurities & 5.0 & 2 \\
$\mu p$ diffusion & 0.5 & 0.5 \\
$\mu + p$ scattering & 3 & 1 \\
$\mu$ pileup veto efficiency & 3 & 2 \\
Analysis methods & 5 & 3 \\
Muon kinetics & 5.8 & 2 \\
EH interference & - & 1 \\
\hline
Total systematic & 10.7 & 6.4 \\
Statistical & 13.7 & 5.3 \\
\hline
 \end{tabular}
 \caption{List of errors in \LS\ for the published data \cite{Andreev:2007wg} and the projected errors for the full statistics.\label{tab:errors}}
 \label{tab:systematics}
 \end{table}


\begin{theacknowledgments}
This work was supported in part by the U.S. National Science Foundation, the U.S. Department of Energy and CRDF, PSI, the Russian Academy of Sciences, and a grant of the President of the Russian Federation (NSH-3057.2006.2). Essential computing resources for the analysis were provided by the National Center for Supercomputing Applications.
\end{theacknowledgments}



\bibliographystyle{aipproc}   

\bibliography{winter}

\end{document}